# The "Jeans Swindle": the end of a myth?

A. I. Ershkovich

*Department of Geophysics and Planetary Sciences*
*Tel Aviv University*

(e-mail: alexer@post.tau.ac.il)

**Abstract.** The Jeans model is shown to be self-consistent, so that the "Jeans Swindle" has never taken place.

Jeans (1902) [1] gave the first quantitative description of fragmentation of infinite uniform self-gravitating gas at rest. This model is called the Jeans model. It is generally believed (see, e.g., [2-5]) to suffer of basic inconsistency, namely: the Jeans model in the equilibrium state does not obey the Poisson equation for the gravitational potential φ

$$\nabla^2 \varphi = 4\pi G \rho \qquad (1)$$

where $G$ is the gravitational constant, ρ is the density. Binney and Tremaine [4-5] even called this inconsistency the "Jeans Swindle". Toward the 100[th] anniversary of the Jeans (1902) famous paper, Kiessling [6] tried to rectify the situation by giving a mathematically rigorous derivation of the Jeans instability criterion: the unperturbed gas is in a static equilibrium if the cosmological constant Λ is introduced. Although Λ tends to zero afterwards such a modified Jeans model differs from the original one. On the other hand, other modified models, e.g. expanding world models (cf. [3] for details) also result in almost the same Jeans instability criterion. Thus, no doubt, the Jeans instability exists, its criterion is quite correct but situation did not change: delegitimation of the Jeans model still continues ([5]).

Let us furnish yet another proof that the Jeans results are correct. The Jeans dispersion relation is

$$\omega^2 = k^2 c_s^2 - \Omega^2 \qquad (2)$$

where $k$ is the wavenumber, $c_s$ is the speed of sound, $\Omega = \sqrt{4\pi G \rho}$ is the Jeans frequency. The instability criterion, naturally, is $\omega^2 < 0$. Equation (2) is similar to the dispersion equation

$$\omega^2 = 3k^2 v_T^2 + \omega_L^2 \qquad (3)$$

which describes Langmuir waves [7], $\omega_L$ is the Langmuir frequency, $v_T$ is the thermal velocity, $\sqrt{3}v_T$ is the speed of electronic sound. This obvious analogy follows also from a comparison between Newton gravity law and Coulomb law (cf. [8]).

Had Jeans drawn attention to this comparison I believe Langmuir waves might have been discovered a quarter of century earlier. Notice that Langmuir oscillations may arise in the system of charged particles of the same sign, say, in the electron gas or beam. Equation (3) does not require the plasma quasineutrality (cf. [8]), so that the similarity with the Jeans equation (2) is obvious. Nevertheless, the Langmuir equation (3), in contrast to (2) has never been associated with a "swindle". Kinetic treatment of the Jeans instability [9-12] shows that the Jeans instability is associated with the collisionless Landau damping.

Thus, since 1902, the Jeans instability became an integral part of physics, astrophysics, cosmology and plasma physics. So why the Jeans model is still associated with a "swindle"? I think, this strange situation depends on attitude (and requirements) towards a physical model. Successful physical model, of course, has to be self-consistent but not necessarily realistic. After all, the last word belongs to experiment. Models of ideal fluid or perfect gas are very fruitful but not realistic. Maxwell demon and Schroedinger cat do not exist either. The Jeans model is yet another example of extremely fruitful and self-consistent but not realistic models.

Finally, let us show that the Jeans model is self-consistent. Indeed, equation of hydrostatics (for gas at rest) is

$$\nabla p_0 = \rho_0 \mathbf{g} \qquad (4)$$

(index "zero" refers to equilibrium). Taking $p_0 = Const$, $\rho_0 = Const \neq 0$, one may arrive at the conclusion (cf. [3]) that uniform matter cannot be in static equilibrium. Of course, this is correct for a finite system but may be incorrect for infinite one. For the Jeans infinite, homogeneous and isotropic model, equation (4) with $p_0 = Const$, $\rho_0 \neq 0$ yields $\mathbf{g} = 0$. Gravity vanishes. Well, it is not realistic but is quite obvious (and self-consistent). In homogeneous and isotropic gas the resulting gravitational force at any point of infinite space equals zero owing to symmetry reasons. But if there is no gravity in equilibrium the Poisson equation (1) (which describes the gravity) becomes irrelevant. The very concept of the gravitational potential is dubious for homogeneous isotropic matter as there is no preferred direction for the vector of force $\nabla \varphi_0$ to point, and for zero force at any point of space this concept becomes meaningless.

Thus, the Jeans model is self-consistent, so that the "Jeans swindle" does not exist and has never taken place.